\documentclass[%
 reprint,
 amsmath,amssymb,
 aps,
]{revtex4-2}
\usepackage{graphicx}% Include figure files
\usepackage{dcolumn}% Align table columns on decimal point
\usepackage{bm}% bold math
\usepackage{orcidlink}
\begin{document}

\preprint{APS/123-QED}

\title{Baumgarte-Shapiro-Shibata-Nakamura formalism in cylindrical coordinates: Brill and Teukolsky waves in both linear and nonlinear regimes}

% \title{Generalized Baumgarte-Shapiro-Shibata-Nakamura formalism in cylindrical coordinates}% Force line breaks with \\

\author{M. A. Alcoforado}
 \email{mariana.alcoforado@uerj.br}
\author{R. F. Aranha \orcidlink{0000-0002-9832-2379}}%
 \email{rafael.aranha@uerj.br}
\author{H. P. de Oliveira \orcidlink{0000-0003-1996-5016}}%
 \email{henrique.oliveira@uerj.br} 
\affiliation{%
Department of Theoretical Physics - Institute of Physics\\
Rio de Janeiro State University - 524 São Francisco Xavier Street - 20550-900\\ 
Maracanã - Rio de Janeiro - RJ - Brazil\\  
}%


\date{\today}% It is always \today, today,

\begin{abstract}
% Following on from our previous work \cite{alcoforado}, we now study the generalized Baumgarte-Shapiro-Shibata-Nakamura formalism described in cylindrical coordinates, extending our code built at Rio de Janeiro State University. This code uses the Galerkin-collocation method and the first numerical applications are related to Brill and Teukolsky gravitational waves. 

In this article we present a numerical code, based on the collocation or pseudospectal method, which integrates the equations of the BSSN formalism in cylindrical coordinates. In order to validate the code, we carried out a series of tests, using three groups of initial data: i) pure gauge evolution; ii) Teukolsky quadrupole solution for low amplitudes and iii) Brill and Teukolsky solutions with higher amplitudes, which accounts for a deviation from the linear regime when compared to the case of low amplitudes. In practically all cases, violations of the Hamiltonian and momentum constraints were analyzed. We also analyze the behavior of the lapse function, which can characterize the collapse of gravitational waves into black holes. Furthermore, all three groups of tests used different computational mesh resolutions and different gauge choices, thus providing a general scan of most of the numerical solutions adopted.

%\begin{description}
%You may use the \texttt{description} environment to structure your %abstract;
%use the optional argument of the \verb+\item+ command to give the %category of each item. 
%\end{description}
\end{abstract}

\maketitle

%\tableofcontents

\section{Introduction}
\indent \par The Baumgarte-Shapiro-Shibata-Nakamura formalism (BSSN)\cite{bssn1,bssn2} is one of the most successful formalisms in Numerical Relativity, as it naturally fits the condition of global hyperbolicity of Einstein's equations for a well-posed Cauchy problem \cite{bruhat}. Despite the success of the method in the application of several astrophysical systems (mainly in obtaining gravitational waveforms and the production of numerical catalogs provided to the LIGO consortium \cite{ligo}), the BSSN formalism only allowed the construction of codes with cartesian coordinate systems. This limitation did not prevent the massive development of the leading scientific articles in the history of this area of research. As main articles, we can highlight the three seminal articles on the coalescence of black holes, including all phases of the collision of the binary system \cite{pretorius, campanelli, baker}. Even with the coordinate system's limitation, the formalism's gauge freedom provides a high diversity of simplifications and dynamical scenarios. And this was also decisive for the plethora of works developed. The {\it moving punctures} method for the dynamical evolution of black hole singularities is also an example of the successful combination of two gauge choices allowed by the method: {\it 1 + log slicing} and the {\it Gamma driving shift condition} \cite{1log,gdrvr}. In addition to the intrinsic issues of formalism, computational methods, and their respective tools have also developed enormously, which can be seen in the computational consortium Einstein Toolkit \cite{etoolkit}, with more than a decade of development and many citations.

%\indent \par Despite the success of the work in Numerical Relativity as described above, it was still important that more coordinate systems were introduced into the BSSN formalism. Brown \cite{brown}, elegantly, introduced the covariance of the three-dimensional sector of the BSSN formalism, allowing the use of any coordinate system and considerably increasing the number of problems to be addressed, always associated with the proposed geometry. For example, spherical coordinates are commonly used for problems related to gravitational collapse with spherical symmetry. This is the simplest choice, as they computationally are represented by the so-called one-dimensional codes (spatially speaking).  In this context, we can highlight the excellent works of Baumgarte {\it et al.} \cite{montero} and Alcubierre and Mendez \cite{alcumendez}. In the first, in addition to obtaining the formalism equations, the authors developed their own code to deal with singular points on the numerical grid. Their code is known in the literature as the PIRK method. With this, they applied it to several systems in order to validate the code, which they achieved with very good precision. In the second (in addition to the excellent pedagogical way of presenting the topic), the authors introduce and analyze the issue of regularizing the origin of the coordinate system. Both works are excellent for introducing the topic to researchers who seek to understand the issue of the BSSN formalism in generalized coordinates and apply it from scratch.
%
\indent \par Despite the success of the work in Numerical Relativity as described above, it was still important that more coordinate systems were introduced into the BSSN formalism. Brown \cite{brown} elegantly introduced the covariance of the three-dimensional sector of the BSSN formalism, allowing the use of any coordinate system and considerably increasing the number of problems to be addressed, always associated with the proposed geometry. For example, spherical coordinates are commonly used for issues related to gravitational collapse with spherical symmetry. In this context, we can highlight the excellent works of Baumgarte {\it et al.} \cite{montero} and Alcubierre and Mendez \cite{alcumendez}. In the first, in addition to obtaining the formalism equations, the authors developed their code to deal with the singular points on the numerical grid. Their code is known in the literature as the PIRK method. They applied this to several systems in order to validate the code, which they achieved with very good precision. In the second (in addition to the excellent pedagogical way of presenting the topic), the authors introduce and analyze the issue of regularizing the origin of the coordinate system. Both works are excellent for introducing the topic to researchers who seek to understand the problem of the BSSN formalism in generalized coordinates and apply it from scratch.

%\indent \par Inspired by the work of Hilditch {\it et al.} \cite{hilditch}, we consider in this work two systems of interest to be analyzed numerically: the spacetimes of Teukolsky \cite{teukolsky} and Brill \cite{brill} which represent gravitational waves. Although they do not represent a specific radiation source, such as black hole binaries or rotating stars, Teukolsky and Brill waves have similarities in the linear regime and serve as an excellent numerical laboratory to explore the BSSN formalism, different gauge choices and numerical methods. Hilditch {\it et al.} explored some characteristics of similarity between the two cases and, surprisingly, verified that, in the linear regime of the Brill waves, the modes beyond the quadrupole case all vanish. This is achieved accordingly with the seed functions adopted by the authors. This brought to the conclusion that the possible differences between Brill and Teukolsky waves can only be seen in the non-linear regime of Einstein's equations.
%

The evolution of vacuum axisymmetric spacetimes is an advanced theoretical laboratory to investigate the collapse of pure gravitational waves and, simultaneously, to produce numerical codes capable of correctly simulating the dynamics of such spacetimes. The two most common gravitational-wave initial data are the Brill wave, whose description is found in Ref. \cite{brill}, and the Teukolsky wave, which consists of using the linearized Teukolsky solution \cite{teukolsky} to solve the constraint equations to generate a nonlinear initial data. Recently, Baumgarte et al. \cite{baum_23} performed careful numerical work with Brill and Teukolsky waves concerning the critical phenomena in axisymmetric collapse that become a long-standing problem in numerical relativity (see the references of \cite{baum_23}).

%\indent \par For this reason, in this work, we study the comparative evolution between Brill and Teukolsky waves within the non-linearity of the field equations, via BSSN formalism in cylindrical coordinates. The choice of this coordinate system is due to the fact that both cases fit into solutions with axial symmetry. Specifically, the Brill solution itself is already described with a cylindrical type coordinate system, as can be seen in typical solutions of Weyl spacetime classes \cite{kramer}, but for the dynamical case. Computationally, we leave spherical symmetry and introduce an additional dimension for solving a two-dimensional problem for the coordinates $\rho$ and $z$.

\indent \par We consider here the nonlinear evolution of axisymmetrical Brill and Teukolsky waves. We set up the BSSN equations in cylindrical coordinates for both systems and implemented a low-cost computational spectral code based on the Galerkin-Collocation method \cite{hpo_er_14,hpo_clem_17,barreto_etal_19}. Due to cylindrical coordinates, we face the $1/\rho,\,1/\rho^2$ terms in the field equations that require a regularity procedure. We establish basis functions that yield a natural regularization \cite{alcoforado}, for instance, imposing that some of the basis functions in $\rho$ behave as $\mathcal{O}(\rho^2)$ or $f(z)+\mathcal{O}(\rho^2)$. Further, we also implemented the spacetime evolution under several gauge conditions, precisely the $1+log$ slicing, the harmonic slicing, the maximal slicing, and the shock-avoiding slicing \cite{alcub_97}.      

%The content of the article will be as follows. In Sec.\ref{sec02} we present the BSSN formalism in curvilinear coordinates. This includes the dynamic equations and the Hamiltonian and momentum constraints, in addition to the possible gauge choices. In Sec.\ref{sec03} we discussed the main properties of Brill and Teukolsky waves, focusing much of the attention on the seed functions which introduce the main physical parameters considered in the initial data for the numerical scheme to be adopted later. In Sec.\ref{sec04}, the numerical setup is presented. As a typical feature of the numerical codes based on spectral collocation methods, the basis functions best adapted to the regularization conditions at the origin and infinity are used, in addition to a small demonstration of how the dynamical equations are written according to the method. All numerical results are displayed in Sec.\ref{sec05}, with an in-depth discussion about the convergence of the method given the different resolution choices of the numerical grid and the different gauge choices. Finally, in Sec.\ref{sec06}, we present a summary of the results and point out the main future developments associated with the current work.

The content of the article will be as follows. In Sec.\ref{sec02}, we present the BSSN formalism in curvilinear coordinates. This includes the dynamic equations, the Hamiltonian and momentum constraints, and the possible gauge choices. In Sec.\ref{sec03}, we discussed the main properties of Brill and Teukolsky waves, focusing much of the attention on the seed functions, which introduce the main physical parameters considered in the initial data for the numerical scheme to be adopted later. In Sec.\ref{sec04}, the numerical setup is presented. As a typical feature of the numerical codes based on the spectral Galerkin Collocation method, the basis functions are adapted to the regularization at the origin, as we have mentioned. Also, the basis functions guarantee the asymptotic flatness condition. We display the numerical results in Sec.\ref{sec05}, with an in-depth discussion about the convergence of the method given the different resolution choices of the numerical grid and the other gauge choices. Finally, in Sec.\ref{sec06}, we summarize the results and point out the main future developments associated with the current work.

\section{ The BSSN formalism in curvilinear coordinates}
\label{sec02}
\indent \par In this section we describe the BSSN formalism in curvilinear coordinates, based on the references \cite{brown,montero}. We start from the ADM axisymmetric line element in the absence of rotation and written in spherical coordinates \cite{numrelbook},
%
% \begin{eqnarray}
% \label{metricadm}
% \nonumber
%  ds^{2} &=& - \alpha^{2} dt^{2}+ e^{-4\phi} (h_{\rho \rho} d\rho^{2} + \rho^{2} h_{\theta \theta} d\theta^{2} + 2h_{\rho z}d\rho dz  +\\
%  &+& h_{zz} dz^{2} ).  
% \end{eqnarray}
%
\begin{eqnarray}
\label{metricadm}
\nonumber
 ds^{2} &=& - \alpha^{2} dt^{2}+ {\gamma}_{r r}(dr + \beta^{r}dt)^2 + 
 {\gamma}_{\theta \theta}d\theta^2 +\\
 \nonumber
 &+& {\gamma}_{\phi \phi}(d\phi + \beta^{\phi}dt)^2.\\
\end{eqnarray}
\noindent Changing the coordinate system from spherical to cylindrical ($\rho = r \sin \phi$ and $z =r\cos \phi$), a 3-metric component $\gamma_{\rho z}$ is induced. Combining this with the fact that, in the BSSN formalism, we include a conformal transformation,
\begin{eqnarray}
\gamma_{ij} = e^{4\phi} \bar{\gamma}_{ij},
\end{eqnarray}
\noindent the line element finally gets the following form,
\begin{eqnarray}
\label{metricadm}
\nonumber
 ds^{2} &=& - \alpha^{2} dt^{2}+ e^{4\phi} [\bar{\gamma}_{\rho \rho}(d\rho + \beta^{\rho}dt)^2 + \bar{\gamma}_{\theta \theta}d\theta^2 +\\
 \nonumber
 &+& \bar{\gamma}_{zz}(dz + \beta^{z}dt)^2 + 2\bar{\gamma}_{\rho z}(d\rho + \beta^{\rho}dt)(dz + \beta^{z}dt)].\\
\end{eqnarray}
\noindent Here, $\bar{\gamma}_{ij}$ are called the conformal $3$-metric components. We also include new variables $h_{ij}$, related to $\bar{\gamma}_{ij}$ by
\begin{eqnarray}
\bar{\gamma}_{ij} =
\begin{pmatrix}
h_{\rho \rho} & 0 & h_{\rho z}\\
0 & \rho^2 h_{\theta \theta} & 0 \\
h_{\rho z} & 0 & h_{zz}
\end{pmatrix}.
\end{eqnarray}
\noindent Also in (\ref{metricadm}), $\beta^{\rho}$ and $\beta^{z}$ are the nonzero cylindrical components of the shift vector, $\beta^{j}$. Then, $\beta^{j}=(\beta^{\rho}, 0 , \beta^{z})$.
%
% \indent \par As discussed in \cite{chandra}, the metric of a sufficiently general spacetime for the dynamical and axisymmetric case is given by (with a positive signature) 
% %
% \begin{eqnarray}
% \nonumber
%  ds^{2} &=& -e^{2\nu} dt^{2} + e^{2\psi} (d\varphi - q_2 dx^2 - q_3 dx^3 -\omega dt)^2 \\
%  &+& e^{2\mu_2} (dx^{2})^2 + e^{2\mu_3} (dx^{3})^2,  
% \end{eqnarray}
% %
% \noindent where $\nu$, $\psi$, $q_2$, $q_3$, $\omega$, $\mu_2$ and $\mu_3$ are functions of $t$, $x^2$ and $x^3$. This is exactly the case of equation (\ref{metricadm}) (considering a additional transformation of $x^2 = x^{2}(\rho,z)$ and $x^3 = x^{3}(\rho,z)$ coordinates). Despite removing two functions ($\omega=0$ and $q_2 = q_{2}(q_3)$), this form of the metric, as we will see in the results obtained, can still be considered sufficiently general in the context of Chandrasekhar \cite{chandra}. Of course, by removing more functions (or its degrees of freedom) from the metric above, it will result in the stationary case and this leads to inconsistencies in the BSSN formalism, as we will point out later. Therefore, care must be taken not to remove the dynamical degrees of freedom from the Einstein equations represented by the equations below. 
%
\indent \par Before presenting the dynamical equations of the generalized BSSN formalism, it is necessary to point out the different definitions of the formalism. First, the rescaled and traceless extrinsic curvature,
\begin{eqnarray}
\tilde{A}_{ij} = e^{-4\phi} (K_{ij} - \frac{1}{3}\gamma_{ij}K).
\end{eqnarray}
\noindent It is important to highlight that we used, via axisymmetry, a specific form of $\tilde{A}_{ij}$,
\begin{eqnarray}
\tilde{A}_{ij} =
\begin{pmatrix}
a_{\rho \rho} & 0 & a_{\rho z}\\
0 & \rho^2 a_{\theta \theta} & 0 \\
a_{\rho z} & 0 & a_{zz}
\end{pmatrix}.
\end{eqnarray}
\noindent A significant step towards a covariant description of the ADM formalism lies in the structure formed by the difference of two connections, given by
\begin{eqnarray}
\label{diffconn}
\Delta\Gamma^{i}_{jk} = \bar{\Gamma}^{i}_{jk} - \mathring{\Gamma}^{i}_{jk},
\end{eqnarray}
\noindent from which we can get a vector
\begin{eqnarray}
\label{Gamvec}
\Delta\Gamma^{i} = \bar{\gamma}^{jk} \Delta\Gamma^{i}_{jk}.
\end{eqnarray}
\noindent In (\ref{diffconn}), the term $\mathring{\Gamma}^{i}_{jk}$ is the connection constructed through the flat metric written in cylindrical coordinates,
\begin{eqnarray}
\mathring{\gamma}_{ij} =
\begin{pmatrix}
1 & 0 & 0\\
0 & \rho^2 & 0 \\
0 & 0 & 1
\end{pmatrix}.
\end{eqnarray}
\noindent It is often advantageous to define a new vector quantity, $\Lambda^{i}$, in order to check the numerical evolution of the vector given by (\ref{Gamvec}),
\begin{eqnarray}
\mathcal{C}^{i} = \Lambda^{i} - \Delta\Gamma^{i} = 0.
\end{eqnarray}
\noindent The components of $\Lambda^{i}$ are given by
\begin{eqnarray}
\Lambda^{i} = (\Lambda^\rho , 0 , \Lambda^z).
\end{eqnarray}
\noindent According to \cite{brown}, there is a freedom to adopt a Lagrangian referential associated with the evolution of the determinant of the metric as,
\begin{eqnarray}
\partial_t \bar{\gamma}=0.
\end{eqnarray}
\noindent This condition is important, as it eliminates the density weight and guarantees the tensor character of the Lie derivative along the shift vector for all quantities of the BSSN formalism.
\indent \par That said, we can list the dynamical equations associated with the variables $\phi$, $\bar{\gamma}_{ij}$, $K$, $\bar{\Lambda}^{i}$ and $\tilde{A}_{ij}$:
\begin{itemize}
\item[i)] {\it{Conformal factor}} 
\begin{eqnarray}
\partial_{t} \phi = \beta^{i}\partial_{i}\phi + \frac{1}{6}\bar{\nabla}_{i}\beta^{i} - \frac{1}{6} \alpha K;
\end{eqnarray}
\item[ii)] {\it{Conformal metric components}} 
\begin{eqnarray}
\nonumber
\partial_t \bar{\gamma}_{ij} &=& \beta^{k}\partial_{k}\bar{\gamma}_{ij} + \bar{\gamma}_{ik}\partial_{j}\beta^{k} +
\bar{\gamma}_{kj}\partial_{i}\beta^{k} +\\
&-&\frac{2}{3}\bar{\gamma}_{ij}\bar{\nabla}_{k}\beta^{k}  - 2\alpha \tilde{A}_{ij};
\end{eqnarray}
\item[iii)] {\it{Trace of the extrinsic curvature}} 
\begin{eqnarray}
\label{eqK}
\nonumber
\partial_t K &=& \beta^{i}\partial_{i} K +\frac{\alpha}{3} K^2 + \alpha \tilde{A}_{ij}\tilde{A}^{ij} - e^{-4\phi} (\bar{\nabla}^2 \alpha +\\
&+& 2 \bar{\nabla}^{i} \alpha \bar{\nabla}_{i} \phi );
\end{eqnarray}
\item[iv)] {\it{Vector from the connection difference}} 
\begin{eqnarray}
\nonumber
\partial_t \bar{\Lambda}^{i} &=&
\bar{\gamma}^{jk} \mathring{\nabla}_{j}\mathring{\nabla}_{k} \beta^{i} + \frac{2}{3} \bar{\Lambda}^{i}\bar{\nabla}_{j}\beta^{j} + \frac{1}{3} \bar{\nabla}^{i}\bar{\nabla}_{j}\beta^{j}+ \\
&+&\beta^{k}\partial_{k}\bar{\Lambda}^{i} -
\bar{\Lambda}^{k}\partial_{k}\beta^{i}
- 2 \tilde{A}^{jk}(\delta^{i}_{j} \partial_{k} \alpha - \\
\nonumber
&-& 6 \alpha \delta^{i}_{j}\partial_{k} \phi - \alpha \Delta\Gamma^{i}_{jk}) - \frac{4}{3} \alpha \bar{\gamma}^{ij} \partial_{j} K; \\
\end{eqnarray}
\item[v)] {\it{Traceless extrinsic curvature rescaled}} 
\begin{eqnarray}
\nonumber
\partial_t \tilde{A}_{ij} &=& 
\beta^{k}\partial_{k}\tilde{A}_{ij} + \tilde{A}_{ik}\partial_{j}\beta^{k} +
\tilde{A}_{kj}\partial_{i}\beta^{k} -\frac{2}{3}\bar{A}_{ij}\bar{\nabla}_{k}\beta^{k}  +\\
\nonumber
&-& 2\alpha \tilde{A}_{ik} \tilde{A}^{k}_{j} + \alpha K \tilde{A}_{ij} + e^{-4\phi} [ - 2\alpha \bar{\nabla}_{i} \bar{\nabla}_{j} \phi +\\ 
\nonumber
&+& 4 \alpha \bar{\nabla}_{i} \phi \bar{\nabla}_{j} \phi + 4 \bar{\nabla}_{(i} \alpha \bar{\nabla}_{j)} \phi - \bar{\nabla}_{i} \bar{\nabla}_{j} \alpha +\\ 
&+& \alpha \bar{R}_{ij}]^{TF},
\end{eqnarray}
\end{itemize}

\noindent where the conformal Ricci tensor components are given by
\begin{eqnarray}
\nonumber
\bar{R}_{ij} &=& -\frac{1}{2} \bar{\gamma}^{kl} \mathring{\nabla}_{k} \mathring{\nabla}_{l} \bar{\gamma}_{ij} + \bar{\gamma}_{k(i} \mathring{\nabla}_{j)} \bar{\Lambda}^{k} + \Delta\Gamma^{k} \Delta\Gamma_{(ij)k} \\
&+& \bar{\gamma}^{kl} (2\Delta \Gamma^{m}_{k(i}\Delta \Gamma_{j)ml} + \Delta \Gamma^{m}_{ik} \Delta \Gamma_{mjl}),
\end{eqnarray}
\noindent and $TF$ means the tracefree part of a specific quantity. Here, $\mathring{\nabla}$ means a covariant derivative with respect to the flat background metric components $\mathring{\gamma}_{ij}$. The generic quantity $\chi_{ij}$ has its tracefree part defined by
\begin{eqnarray}
\chi_{ij}^{TF} \equiv \chi_{ij} - \frac{1}{3}\gamma_{ij} \gamma^{kl}\chi_{kl} = \chi_{ij} - \frac{1}{3}\bar{\gamma}_{ij} \bar{\gamma}^{kl}\chi_{kl} .
\end{eqnarray}
\indent \par In addition to the dynamical equations described above, the formalism requires that the constraint equations must be satisfied on each hypersurface with constant $t$. The first of these two equations is the Hamiltonian constraint which can be given by the following expression, 
\begin{eqnarray}
\nonumber
\mathcal{H} &\equiv& \frac{2}{3}K^2 - \tilde{A}_{ij} \tilde{A}^{ij} + e^{-4\phi}(\bar{R} - 8 \bar{\nabla}^{i}\phi \bar{\nabla}_{i}\phi - 8 \bar{\nabla}^2 \phi).\\
\end{eqnarray}
\indent \par The second equation is given by the momentum constraint. The expression used here in this paper is a variation of the equivalent equation of the original ADM formalism \cite{adm}, considering the divergence of the traceless extrinsic curvature,
\begin{eqnarray}
\nabla_{j}A^{ij} = e^{-4\phi}(\bar{\nabla}_{j}\tilde{A}^{ij} + 6\tilde{A}^{ij} \partial_{j}\phi).
\end{eqnarray}
\noindent Then, the momentum constraint equation can be displayed as
\begin{eqnarray}
\mathcal{M}^{i} = e^{-4\phi}(\bar{\nabla}_{j}\tilde{A}^{ij} + 6\tilde{A}^{ij} \partial_{j}\phi - \frac{2}{3}\bar{\gamma}^{ij}\partial_{j}K).
\end{eqnarray}
\noindent Another important point in the development of the BSSN formalism is the gauge choice, as there is a freedom to choose how the lapse function and the shift vector evolve. This can be done since there are infinite ways on slicing the manifold as a 3+1 spacetime with a Cauchy maximal development. A important gauge expression chosen for the evolution of the lapse function is given by the {\it 1+log} condition,
\begin{eqnarray}
(\partial_t -\beta^{j}\partial_{j}) \alpha = - 2f(\alpha) K,
\end{eqnarray}
\noindent where $f(\alpha)$ is a function of the lapse function, $\alpha$. Tests on different gauge expressions for $f(\alpha)$ are explored in Section \ref{sec05}. Also, we can choose the {\it Gamma driver} condition based on the time evolution of the shift vector,
\begin{eqnarray}
(\partial_t -\beta^{j}\partial_{j}) \beta^{i} = \frac{3}{4}B^{i},
\end{eqnarray}
\noindent where the auxiliary vector $B^{i}$ is evolved through
\begin{eqnarray}
(\partial_t -\beta^{j}\partial_{j}) B^{i} = \partial_{t}\bar{\Lambda}^{i} - \eta B^{i}.
\end{eqnarray}
\noindent The directional derivatives $\beta^{j}\partial_{j}$ are known as advective terms and can be ignored depending on the numerical setup. The constant $\eta$ is of the order of $1/2M$, where $M$ is the total mass of the spacetime. Another gauge condition to be used in this work is the maximal slicing, where the trace of the extrinsic curvature is zero initially and at all subsequent times,
\begin{eqnarray}
K=0=\partial_t K.
\end{eqnarray}
\noindent In this gauge, the evolution equation for $K$ (eq.\ref{eqK}) fixes the spatial variations of the lapse function.
\indent \par Now we have the entire set of equations necessary to write them in any curvilinear coordinate system. In Appendix \ref{app} we show an example of how to obtain the explicit equations for the case of cylindrical coordinate system.

\section{Brill and Teukolsky waves}
\label{sec03}
\indent \par There are some solutions in the literature of General Relativity which represent gravitational waves, both in the linear regime and in the nonlinear regime. From these solutions, we will consider (as initial data) two important cases: the Teukolsky and Brill waves \cite{teukolsky,brill}. While Teukolsky waves represent linearized gravitational waves, Brill waves are classified as vacuum nonlinear waves. It is also important to mention that these solutions are not related to specific sources of gravitational radiation, such as black hole binaries or rotating neutron stars. These solutions simply represent gravitational waves propagating in a vacuum. In the next two subsections we will present the main characteristics of these solutions and the context in which they will serve as initial data to be numerically integrated later. 

\subsection{Teukolsky Waves}
\indent \par Considering spherical coordinates, it is convenient to express the two degrees of freedom of the gravitational waves in terms of the polar and axial modes. A polar mode $(l,m)$ has parity $(-1)^{l}$ under the space inversion $(\theta, \phi) \longrightarrow (\pi - \theta, \phi + \pi) $ and an axial mode has parity $(-1)^{l+1}$ under the same inversion. The quadrupole mode $(l=2)$ represents the dominant mode for several sources of gravitational waves. The Teukolsky solution element for axial and quadrupolar linear gravitational waves is given by the following line element \cite{teukolsky},
\begin{widetext}
\begin{eqnarray}
\label{teukolskymetric}
\nonumber
ds^2 &=& -dt^2 + (1+Af_{rr})dr^2 + (2Bf_{r\theta})rdrd\theta + (2Bf_{r\phi})r\sin \theta dr d\phi + \left(1 + Cf_{\theta \theta}^{(1)} + Af_{\theta \theta}^{(2)}\right)r^2 d\theta ^2 + \\
&+& [2(A-2C)f_{\theta \phi}]r^2 \sin \theta d\theta d\phi + \left(1 + Cf_{\phi \phi}^{(1)} + Af_{\phi \phi}^{(2)}\right)r^2 d\phi ^2
\end{eqnarray}
\end{widetext}
\noindent The coefficients $A$, $B$ and $C$ are constructed with an arbitrary seed function $F(x)$, where $x = t-r$ for an outgoing solution and $x=t+r$ for an ingoing solution. Generally, we can write 
\begin{eqnarray}
F = F_{1}(t-r) + F_{2}(t+r).
\end{eqnarray}
\noindent Defining
\begin{eqnarray}
F^{(n)} \equiv \left[ \frac{d^{n} F_{1}}{dx^{n}} \right]_{x=t-r} + (-1)^{n}\left[ \frac{d^{n} F_{2}}{dx^{n}} \right]_{x=t+r},
\end{eqnarray}
\noindent we can write the coefficients as
\begin{eqnarray}
\nonumber
A &=& 3 \left[ \frac{F^{(2)}}{r^{3}} + 3\frac{F^{(1)}}{r^{4}} + 3\frac{F}{r^{5}}\right],\\
\nonumber
B &=& - \left[ \frac{F^{(3)}}{r^{3}} + 3\frac{F^{(2)}}{r^{3}} + 6\frac{F^{(1)}}{r^{4}} + 6\frac{F}{r^{5}}     \right],\\
\nonumber
C &=& \frac{1}{4} \left[ \frac{F^{(4)}}{r} + 2\frac{F^{(3)}}{r^{2}} + 9\frac{F^{(2)}}{r^{3}} + 21\frac{F^{(1)}}{r^{4}} + 21\frac{F}{r^{5}} \right].\\
\end{eqnarray}
\noindent The angular functions $f_{ij}$ in the line element depend on the axial index $m$. Since it is related to the polar $l$ as $(2l+1)$ quantities, we have $m = \pm 2, \pm 1, 0$. As we will analyze numerical evolutions with axial symmetry, we can only consider solutions with $m=0$. To this end, the spherical Teukolsky functions $f_{ij}$ are described by \cite{teukolsky}
\begin{eqnarray}
%\nonumber
%f_{rr} &=& 2-3\sin^2 \theta ,\\
%\nonumber
%f_{r\theta} &=& -3\sin \theta \cos \theta ,\\
%\nonumber
%f_{r\phi} &=& 0,\\
%\nonumber
%f_{\theta \theta}^{(1)} &=& 3\sin^2 \theta ,\\
%\nonumber
%f_{\theta \theta}^{(2)} &=& -1,\\
%\nonumber
%f_{\theta \phi} &=& 0,\\
%\nonumber
%f_{\phi \phi}^{(1)} &=& - 3\sin^2 \theta \\
%f_{\phi \phi}^{(2)} &=& 3\sin^2 \theta - 1.
%
f_{rr} &=& 2-3\sin^2 \theta,\,f_{r\theta} = -3\sin \theta \cos \theta, \nonumber\\
\nonumber \\
f_{r\phi} &=& 0,\,f_{\theta \theta}^{(1)} = 3\sin^2 \theta,\, f_{\theta \theta}^{(2)} = -1, \nonumber\\
\nonumber \\
f_{\theta \phi} &=& 0,\,f_{\phi \phi}^{(1)} = - 3\sin^2 \theta,\,f_{\phi \phi}^{(2)} = 3\sin^2 \theta - 1.
\end{eqnarray}
\indent \par The initial data are going to be represented by a specific seed function constructed as a superposition of gaussian terms as
\begin{eqnarray}
\label{seedteukolsky}
%\nonumber
%F(0,r) = A_0 r \left[ e^{-(r/\lambda)^2} - e^{(r/\lambda)^2}  \right].\\
F_1(u) = A_0 \frac{u}{\lambda} \left[ e^{-((u-r_0)/\lambda)^2} + e^{-((u+r_0)/\lambda)^2}  \right],
\end{eqnarray}

\noindent where $u=t-r$. Also, $\lambda$ and $r_0$ are the width and the center of the gravitational wave, respectively. Here, we choose $F_2(v)=-F_1(u)$. 

A proper initial data constructed with the Teukolsky analytical solution requires the determination of the conformal factor $\psi$ such that we express the three-dimensional line element, $dl^2$, as

\begin{equation}
dl^2 = \psi^4 d\bar{l}^2,
\end{equation}

\noindent where we extracted the three-dimensional line element $d\bar{l}^2$ from (\ref{teukolskymetric}). It is important to point out that, here, we do not use the conformal factor $\phi = \ln{\psi}$ to solve the initial data numerically. This fact occurs since the resulting equation for the Hamiltonian constraint is linear, which does not occur in the case of using the BSSN function, $\phi$. In this way, we obtain the conformal factor $\psi$ after solving the Hamiltonian constraint that is given by

\begin{eqnarray}
\label{teukolskycons}
\mathcal{H} &=& W_{rr}\frac{\partial^2 \psi}{\partial r^2}+W_{r\theta}\frac{\partial^2 \psi}{\partial r \partial \theta}+W_{\theta \theta}\frac{\partial^2 \psi}{\partial \theta^2}+W_{r}\frac{\partial \psi}{\partial r}+ \nonumber \\
\nonumber \\
&&+W_{\theta}\frac{\partial \psi}{\partial \theta}- R\psi = 0.
\end{eqnarray}

%\noindent where $R$ is the three-dimensional Ricci scalar calculated with the exact Teukolsky solution. We have assumed time-symmetric data in which $K_{ij}=0$. The factors $B_{rr}, B_{r \theta},..$ depend on the exact solution. When evolving gravitational wave data constructed with the Teukolsky waves in Section V, we solved the above equation in cylindrical coordinates

\noindent We have assumed $K_{ij}=0$ for the time-symmetric data. The three dimensional Ricci scalar $R$ and the factors $W_{rr}, W_{r \theta}$, etc., depend on the Teukolsky exact solution. When evolving gravitational wave data constructed with the Teukolsky waves in Section V, we have to ensure that the above equation is satisfied on each future hypersurface $t= constant$ (within numerical tolerance).

\subsection{Brill Waves}
\indent \par Brill waves are defined as nonlinear, axisymmetric gravitational waves in vacuum spacetimes which admit a moment of time symmetry \cite{brill}. Its three dimensional line element $dl^2$ can be given by %Such waves are represented by the spatial metric in cylindrical coordinates

\begin{eqnarray}
dl^2 = \psi^{4}[e^{q}(d\rho ^2 +dz^2) + \rho^2 d\phi ^2],
\end{eqnarray}

\noindent were $q(\rho, z)$ is an arbitrary and axisymmetric seed function which introduces a deviation from conformal flatness and that can be considered a measure of the gravitational wave amplitude. %The Hamiltonian constraint shows us that
%
%\begin{eqnarray}
%\label{brillconst}
%\mathcal{H} = \nabla^2 \psi + \frac{\psi}{8} \left[ \frac{\partial^2 q}{\partial \rho^2} + \frac{\partial^2 q}{\partial z^2} \right] = 0. 
%\end{eqnarray}
%
%\noindent The seed function for the Brill wave are subject to the following regularity and asymptotic restrictions,
\noindent The seed functions for the Brill wave have to satisfy the following regularity and asymptotic restrictions,
%
%\begin{eqnarray}
%\nonumber
%q = 0 ~ , ~ \textnormal{for} ~ \rho = 0 ~ ,\\
%\nonumber
%\partial_\rho q = 0 ~ , ~ \textnormal{for} ~ \rho = 0 ~ ,\\
%\nonumber
%\partial_z q = 0 ~ , ~ \textnormal{for} ~ z = 0 ~ ,\\
%q \sim  (\rho^2 + z^2)^{-a/2} ~ , ~ a \geq 2 ~ .
%\end{eqnarray}
%
\begin{eqnarray}
\nonumber
& & q = 0, ~ \partial_\rho q = 0,~ \textnormal{for} ~ \rho = 0 ~ ,\\
%\partial_\rho q = 0 ~ , ~ \textnormal{for} ~ \rho = 0 ~ ,\\
\nonumber
& & \partial_z q = 0 ~ , ~ \textnormal{for} ~ z = 0 ~ ,\\
%\nonumber
& & q \sim  (\rho^2 + z^2)^{-a/2} ~ , ~ a \geq 2 ~ .
\end{eqnarray}

%\noindent We can also use a seed function \cite{eppley} in which a initial data for $\psi$ is achieved after integrating equation (\ref{brillconst}) numerically,

There are two commonly used families of seed functions. The first one is due to Eppley \cite{eppley} and given by 

\begin{eqnarray}
\label{eppley}
q(\rho,z) =\frac{A_0 \rho^2}{1+[(\rho^2 + z^2)^{1/2}/\lambda]^{n}},
\end{eqnarray}
\noindent where $n \geq 4$. The second seed function is a Gaussian distribution proposed by Holz, Miller, Wakano and Wheeler \cite{holz}, %Another seed function \cite{holz}, which will be considered in Section \ref{sec5} is a gaussian distribution
\begin{eqnarray}
\label{holz}
q(\rho,z) =A_0\left(\frac{\rho}{\lambda}\right)^2 e^{-[(\rho -\rho_0)^2 - z^2]/\lambda^2}.
\end{eqnarray}
%
%\noindent where $n \geq 4$, Here, $\sigma$ and $\rho_0$ are the width and the center of the gaussian. 

\noindent In both cases, $A_0$ is the initial amplitude, and the parameters $\lambda$ and $\rho_0$ indicate the width and the center of the Brill wave respectively.

The initial data representing Brill waves is completed with the determination of the conformal factor $\psi(\rho,z)$ after solving the Hamiltonian constraint given by 

\begin{eqnarray}
\label{brillconst}
\nonumber
\mathcal{H} = \frac{\partial^2 \psi}{\partial \rho^2}+\frac{1}{\rho}\frac{\partial \psi}{\partial \rho} + \frac{\partial^2 \psi}{\partial z^2} + \frac{\psi}{8} \left( \frac{\partial^2 q}{\partial \rho^2} + \frac{\partial^2 q}{\partial z^2} \right) = 0.\\ 
\end{eqnarray}

%%%%%%%%%%%%%%%%%%%%%%%%%%
\section{Numerical Setup}%
\label{sec04}
%%%%%%%%%%%%%%%%%%%%%%%%%%

We describe the numerical algorithm based on the Galerkin-Collocation method used to integrate the BSSN equations. The numerical code is an extension of the $\mathrm{RIO ~Code}$ \cite{alcoforado} for the use of cylindrical coordinates and establishing a 2D spatial integration. It is necessary first to establish the boundary conditions the metric functions must satisfy in order to assure the regularity of the spacetime near the origin and its asymptotic flatness character. For the latter condition, it is necessary that $r \sim \sqrt{\rho^2+z^2}$ as  $r \rightarrow \infty$. Then we have
\begin{eqnarray}
&\{ \alpha,\,h_{\rho \rho},\,h_{\theta \theta},\,h_{zz}\,\, \}\rightarrow 1&,
\nonumber \\
&\{ K,\,\phi,\,h_{\rho z},\,a_{\rho \rho},\,a_{\theta \theta},\,\Lambda^{\rho}\,\,\Lambda^{z},\,\beta^{\rho},\,\beta^{z} 
  \} \rightarrow 0&.
\end{eqnarray} 

The regularity conditions near the origin reflect the dependence on $\rho$ since $1/\rho$ and $1/\rho^2$ terms appear in the field equation. Then, it follows that 
\begin{eqnarray}
\alpha,\,h_{\rho \rho},\,h_{\theta \theta},\,h_{zz},\,K,\,\phi,\,a_{\rho \rho},\,a_{\theta \theta},\,\Lambda^z,\,\beta^{z}
\end{eqnarray}

\noindent are all even functions of $\rho$, and 
\begin{eqnarray}
h_{\rho z},\,a_{\rho z},\,\Lambda^\rho,\,\beta^{\rho}
\end{eqnarray}

\noindent are odd functions of $\rho$. In addition, we have the following difference relations,
\begin{eqnarray}
\label{diffrel}
\nonumber
(h_{\rho \rho}-h_{\theta \theta})_{\rho=0}&=&0, \\
\nonumber \\
(a_{\rho \rho}-a_{\theta \theta})_{\rho=0}&=&0.
\end{eqnarray}

\noindent For this reason, it is appropriate to introduce new field variables $\bar{h}_{\rho \rho},\,\bar{h}_{\theta \theta}$ and $\bar{a}_{\rho \rho},\,\bar{a}_{\theta \theta}$ by

\begin{eqnarray}
\bar{h}_{\rho \rho} = \frac{1}{2}(h_{\rho \rho}+h_{\theta \theta}),\quad \bar{a}_{\rho \rho} = \frac{1}{2}(a_{\rho \rho}+a_{\theta \theta}), \\
\nonumber \\
\bar{h}_{\theta \theta} = \frac{1}{2}(h_{\rho \rho}-h_{\theta \theta}),\quad \bar{a}_{\theta \theta} = \frac{1}{2}(a_{\rho \rho}-a_{\theta \theta}). 
\end{eqnarray}

\noindent The relations (\ref{diffrel}) are satisfied provided that $\bar{h}_{\rho \rho} = \bar{h}^{(0)}_{\rho \rho}(z) + \mathcal{O}(\rho^2)$ and $\bar{h}_{\theta \theta} = \mathcal{O}(\rho^2)$ near $\rho=0$ and the same behavior holds for $\bar{a}_{\rho \rho}$ and $\bar{a}_{\theta \theta}$. 

Concerning the dependence on $z$, the functions $h_{\rho z},\,a_{\rho z},\,\Lambda^z$ and  $\beta^z$ are odd functions of $z$, while all remaining functions are even functions of the same coordinate. All aspects relating to the issue of regularization can be seen in \cite{alcumendez}.

Based on the boundary and regularity conditions, the most convenient basis functions we shall adopt are the even and odd sines, $SB_{2j}(\rho)$, $SB_{2j+1}(\rho)$ or $SB_{2j}(z)$, $SB_{2j+1}(z)$, as well as linear combinations of them if necessary. These basis functions are defined by \cite{boyd},

\begin{equation}
SB_k(\rho) = \sin\left[(k+1)\,\mathrm{arccot}\left(\frac{\rho}{L_\rho}\right)\right],
\end{equation} 

\noindent where $L_\rho$ is map parameter, and we obtain the basis $SB_k(z)$ replacing $\rho$ by $z$ and $L_\rho$ by $L_z$. We have employed for the first time these basis functions in connection with the initial data problem in numerical relativity \cite{id_hp}, and also for a spectral code in order to evolve the generalized BSSN equations in spherical symmetry \cite{alcoforado}. Khirnov and Ledvinka \cite{khirnov_ledvinka} considered these basis functions in a quasi-maximal slicing spectral code concerning the evolution of Brill waves.

The spectral approximations of the metric functions consist of a series expansion of the product $SB_k(\rho)\, SB_j(z)$, considering the parity of the metric functions as a guide. For instance, the lapse becomes 

\begin{equation}
\alpha(t,\rho,z)=1+\sum_{k=0}^{N_\rho}\sum_{j=0}^{N_z}\,\hat{\alpha}_{kj}(t)SB_{2k}(\rho)SB_{2j}(z),
\end{equation}

\noindent where $N_\rho$ and $N_z$ are the truncation orders that dictate the number of unknown spectral coefficients $\hat{\alpha}_{kj}(t)$. We establish similar approximations for the metric functions $\bar{h}_{\rho \rho}$ and $h_{zz}$. Likewise, the expansions for the functions $\phi,\, K$ and $\bar{a}_{\rho \rho}$ have the same form of $\alpha-1$ but naturally with distinct spectral coefficients. The spectral approximation for $\bar{h}_{\theta \theta}$ is
  
\begin{equation}
%\bar{h}_{\theta \theta}(t,\rho,z)=\frac{1}{2}\sum_{k,j=0}^{N_\rho, N_z}\,\hat{h}^{(\theta)}_{kj}(t)\left(SB_{2k+2}(\rho)-SB_{2k}(\rho)\right)SB_{2j}(z)
\bar{h}_{\theta \theta}(t,\rho,z)=\sum_{k=0}^{N_\rho}\sum_{j=0}^{N_z}\,\hat{h}^{(\theta)}_{kj}(t)\Psi_{2k}(\rho)SB_{2j}(z)
\end{equation}

\noindent where $\Psi_{2k}(\rho) \equiv \frac{1}{2}\,\left(SB_{2k+2}(\rho)-SB_{2k}(\rho)\right)$ behaves as $\mathcal{O}(\rho^2)$ near the origin as expected. We have established a similar spectral approximation for $\bar{a}_{\theta \theta}$.

The spectral approximations for $h_{\rho z}$ and $a_{\rho z}$ are also similar and share the same basis functions, $SB_{2k+1}(\rho) SB_{2j+1}(z)$. Finally, for the components of the connection and shift vectors $\{ \Lambda^\rho,\,\beta^{\rho} \}$ and $\{ \Lambda^z,\,\beta^{z} \}$, the basis functions are $SB_{2k+1}(\rho) SB_{2j}(z)$ and $SB_{2k}(\rho) SB_{2j+1}(z)$, respectively.

The next step is to obtain the system of ordinary differential equations that approximate the BSSN equations. Therefore, we need to choose a suitable set of $(N_\rho+1)(N_z+1)$ collocation points denoted by $(\rho_l,z_m)$. These points are images of the collocation points defined in the computational domain $-1 \leq x \leq 1$ and $-1 \leq y \leq 1$ under the maps

\begin{eqnarray}
\label{transfxy}
\nonumber
\rho_l&=&\frac{L_\rho y_l}{\sqrt{1-y_l^2}}, \\
\nonumber \\
z_m&=&\frac{L_z x_m}{\sqrt{1-x_m^2}},
\end{eqnarray}

\noindent where $l=0,1,..,2N_{\rho}+2$ and $m=0,1,..,2N_z+2$. However, we will consider only the positive $(N_z+1)$ values of $z_m$ due to the parity of functions about the origin and the positive $(N_{\rho}+1)$ values of $\rho_k$ as the definition of this coordinate requires.

With the spectral approximations of all metric functions, we obtain the residual equations and impose that these vanish at the collocation points $(\rho_l,z_m)$. When we replace the metric functions with their corresponding approximations, we obtain the residual equations related to the evolution equations. For the sake of simplicity, let us consider the evolution equation for the conformal factor $\phi(t,\rho,z)$ in the absence of the shift vector. The residual equation evaluated at each collocation point becomes

\begin{equation}
\mathrm{Res}_{\phi}(t,\rho_l,z_m) = (\phi_{,t})_{lm}+\frac{1}{6} \alpha_{lm} K_{lm} = 0,
\end{equation}

\noindent for all $l=0,1,..,N_{\rho}$, $m=0,1,..,N_{z}$. Thus, we have now a set of $(N_{\rho}+1)(N_{z}+1)$ equations to be evolved from a set of initial data as discussed in the former sections. The number of equations is the same as the numerical values of $\phi$ at the collocation points and starting from the substitution of the initial values of $\alpha$ and $K$. We have implemented the same procedure to transform the BSSN field equations into a system of first-order ordinary differential equations for all other metric functions. 

In establishing the maximal slicing version of the spectral code as mentioned, the equation $K_{,t} =0$ becomes an elliptic-type equation to update the lapse at each spatial slice. By imposing the vanishing of the corresponding residual equation at the collocation points, we ended up with a linear algebraic system for the values $\alpha_{lm}(t)$ (or the modes $\hat{\alpha}_{kj}(t)$) which are solved at each stage of the integration.

We evolve the whole spacetime starting from an initial configuration translated into the values of the relevant metric functions at the collocation points. We integrate afterwards the resulting dynamical system through a Cash-Karp adaptive algorithm \cite{numrecipes}. 

%%%%%%%%%%%%%%%%%%%%%%%%%%%%
\section{Numerical Results}
\label{sec05}%
%%%%%%%%%%%%%%%%%%%%%%%%%%%%
\begin{figure}[h]
	\includegraphics[width=7.5cm,height=5.5cm]{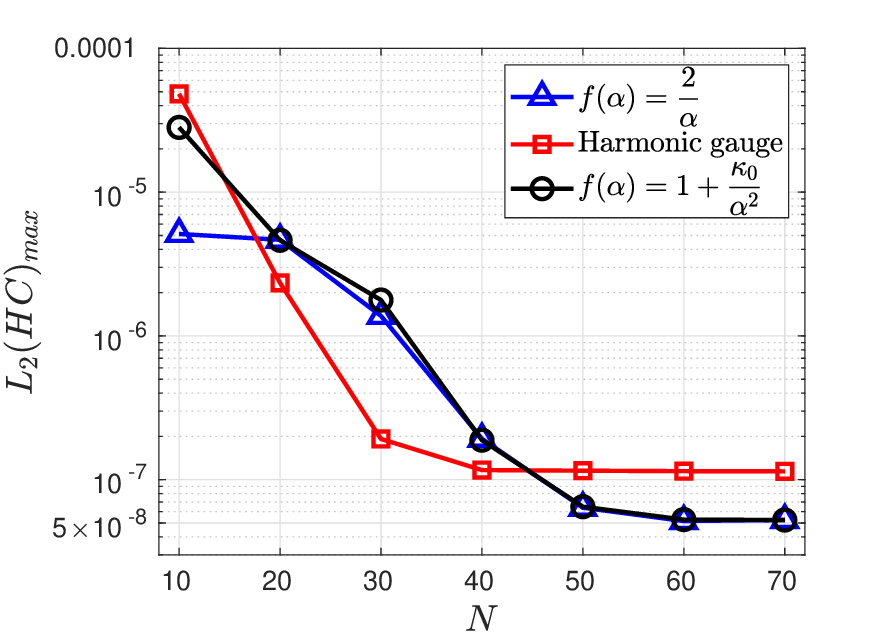}
	\caption{Exponential decay of the maximum values of $L_2$-norm of the initial Hamiltonian constraint for the {\it 1+log} (blue triangles), the harmonic (red squares), and the Alcubierre's shock avoiding (black circles) gauges with $\kappa_0=2$. In the horizontal axis, we have that $N=N_\rho=N_z$, and the map parameters are $L_\rho=L_z=5$.}
\end{figure}	

\subsection{Evolution of pure gauge initial data}

The first numerical test is a standard code convergence of a pure gauge evolution, taking the Hamiltonian and momentum constraint violations as a measure of error. The initial data consists of a gauge perturbation of the Minkowski spacetime with the initial lapse given by

\begin{equation}
\alpha(0,\rho,z) = 1+A_0 \mathrm{e}^{-(\rho^2+z^2)/\sigma},
\end{equation}

\noindent where we have chosen the initial amplitude $A_0=0.01$ and $\sigma=1$. The remaining initial expressions are

\begin{eqnarray}
\bar{h}_{\rho \rho}(0,\rho,z) = h_{zz}(0,\rho,z)=1, 
\end{eqnarray}

\noindent and the remaining functions vanish initially. Here, we have considered the absence of the shift vector, meaning that $\beta^{\rho}=\beta^{z}=0$ at all times.

%\begin{figure}[h]
	%\includegraphics[width=7.5cm,height=5.5cm]{L2HC}
	%\caption{Exponential decay of the maximum values of $L_2$-norm of the Hamiltonian constraint for the $1+log$ (blue triangles), the harmonic (red squares), and the Alcubierre's shock avoiding (black circles) gauges with $\kappa_0=2$. In the horizontal axis, we have that $N=N_\rho=N_z$, and the map parameters are $L_\rho=L_z=5$.}
%\end{figure}	

We have tested the convergence of the Hamiltonian and momentum constraint violations by selecting the corresponding maximum values of their $L_2$-norms\footnote{The standard Hilbert space associated with square-integrable functions is called $L^2[a,b]$ in the literature although we use the $L_2$ notation. In the case of functions of two variables $\rho$ and $z$, the total space is given by $L^2[a_{\rho},b_{\rho}] \otimes L^2[a_{z },b_{z}]$. The functions in this product space are represented by $f(\rho,z)\equiv \left< \rho,z | f\right>$} denoted by $L_2(HC)$, $L_2(MC_\rho)$ and $L_2(MC_z)$ as we increase the truncation orders $N_\rho$ and $N_z$. The $L_2$ norm is defined by

\begin{eqnarray}
\label{normL2}
||f||_{2}=\int_{-\infty}^{\infty} \int_{0}^{\infty}|f(\rho,z)|^2 \rho ~d\rho ~dz. 
\end{eqnarray}

\noindent With transformations (\ref{transfxy}), it is possible to determine the collocation points for $\rho$ and $z$ from the values of $x \in [-1,1]$ and $y \in [-1,1 ]$. However, as mentioned previously, we only consider the computational mesh with $(N_{\rho}+1) \times(N_{z}+1)$ collocation points, all in the positive branch. In this way, both the auxiliary coordinates will have their numerical mesh considered in the interval $[0,1]$. Consequently, the integral (\ref{normL2}), for the error of the Hamiltonian and momentum constraints, will be calculated numerically through the Gauss-Legendre quadrature only inside the positive branch of the induced mesh of $\rho$ and $z$.
\indent \par We further considered the influence of distinct gauge choices on the error decay: the {\it 1+log} gauge, the harmonic gauge, and the called {\it Alcubierre shock avoiding} gauge \cite{alcub_97}. These gauges are characterized  by the following choices of the function $f(\alpha)$:

\begin{equation}
f(\alpha)=\frac{2}{\alpha},
\end{equation}   

\noindent for the {\it 1+log} gauge, and

\begin{equation}
f(\alpha)=1+\frac{\kappa_0}{\alpha^2},
\end{equation}   

\noindent for the harmonic gauge for $\kappa_0=0$, and the Alcubierre's shock avoiding gauge for $\kappa_0 \neq 0$.

\begin{figure}[h]
\includegraphics[width=7.5cm,height=5.5cm]{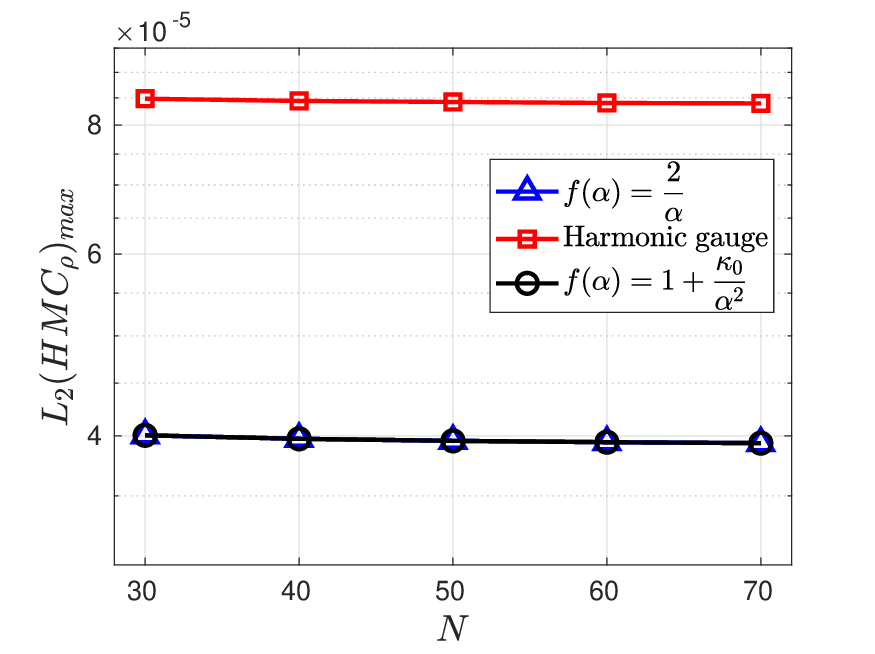}
\includegraphics[width=7.5cm,height=5.5cm]{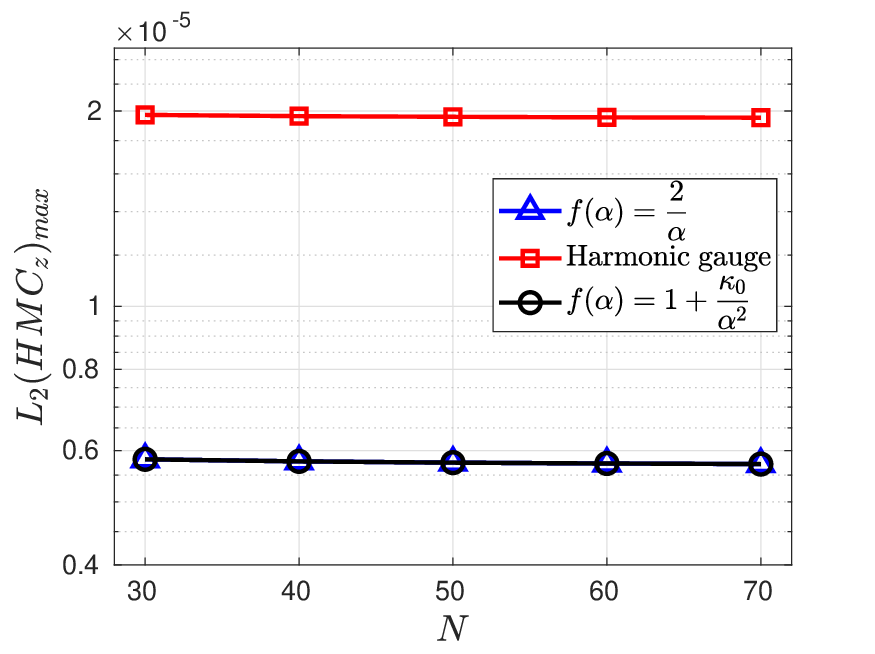}
	\caption{Behavior of the maximum of values of $L_2(MC_\rho)$ and $L_2(MC_z)$ for the {\it 1+log} (blue triangles), the harmonic (red squares), and the Alcubierre shock avoiding (black circles, for $\kappa_0=2$) gauges. We have adopted $N=N_\rho=N_z$, and the map parameters are $L_\rho=L_z=5$. }
\end{figure}	

Fig. 1 depicts the exponential decay of the maximal $L_2(HC)$ for the {\it 1+log} slicing (blue triangles), harmonic gauge (red squares), and the shock avoiding gauge (black circles) with $\kappa_0=2$. We have set the truncation orders equal, $N_\rho=N_z=N$, and the map parameters $L_\rho=L_z=5$. Notably, the {\it 1+log} and the shock-avoiding gauges produce similar error decays and are better than the outcomes due to the harmonic gauge for $N \geq 50$. Concerning the maximal values of $L_2(MC_\rho)$ and $L_2(MC_z)$, we noticed from the plots of Fig. 2 that there is no error decay, although the maximal error is relatively small. However, the {\it 1+log} and the shock-avoiding gauges perform better than the harmonic gauge. 

\begin{figure}[h]
\includegraphics[width=7.5cm,height=5.5cm]{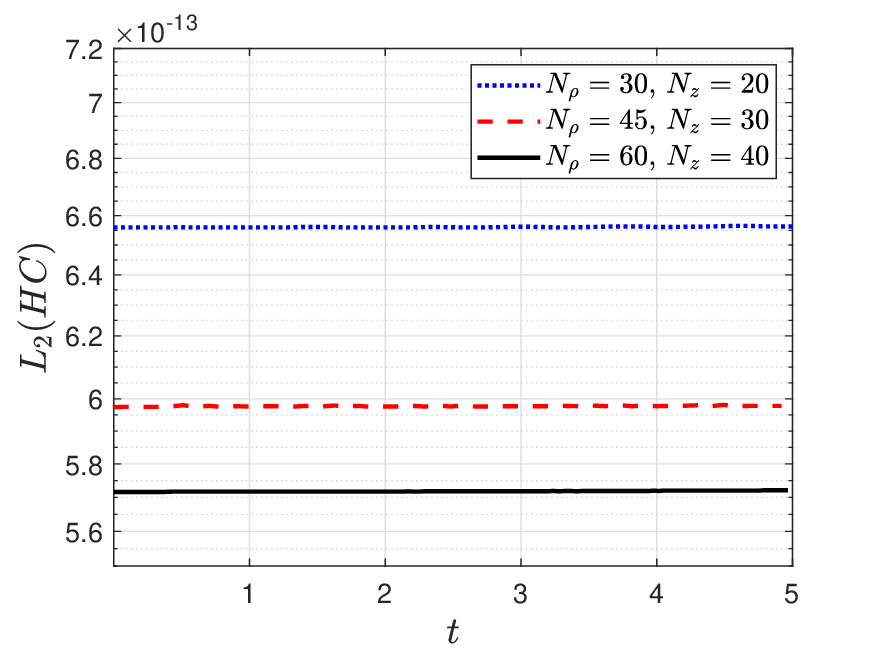}
\caption{Evolution of the violation of the Hamiltonian constraint $L_2(HC)$ for distinct resolutions, $N_\rho=30,45$ and $60$. The initial data is the axisymmetric Teukolsky solution with amplitude $A_0=10^{-7}$. Here, we have used the shock avoiding gauge with $\kappa_0=2$ and $L_\rho=L_z=5$.}
\end{figure}

\begin{figure}[htb]
\includegraphics[width=7.5cm,height=5.5cm]{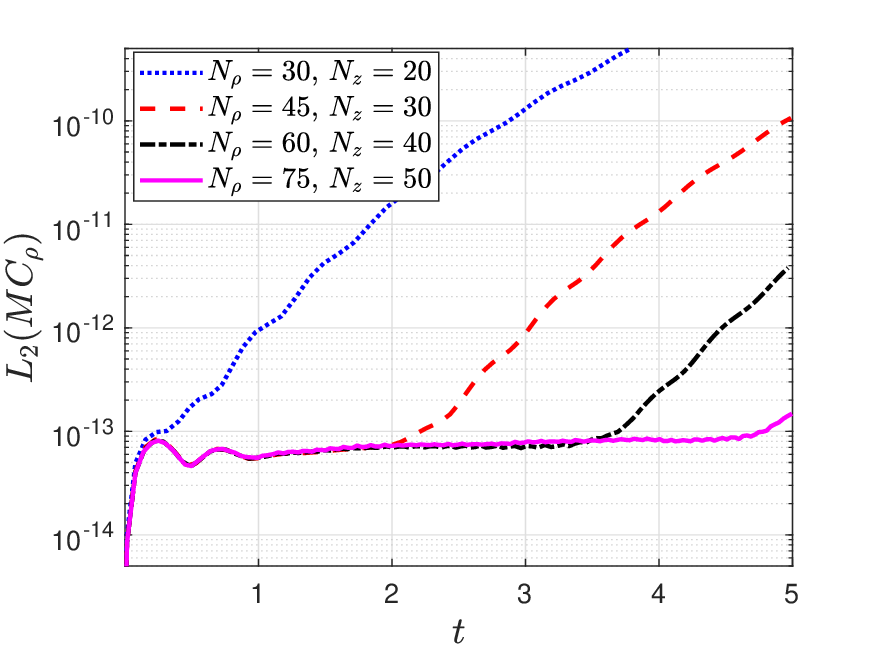}%{L2MC1_TW}\\
\\
\includegraphics[width=7.5cm,height=5.5cm]{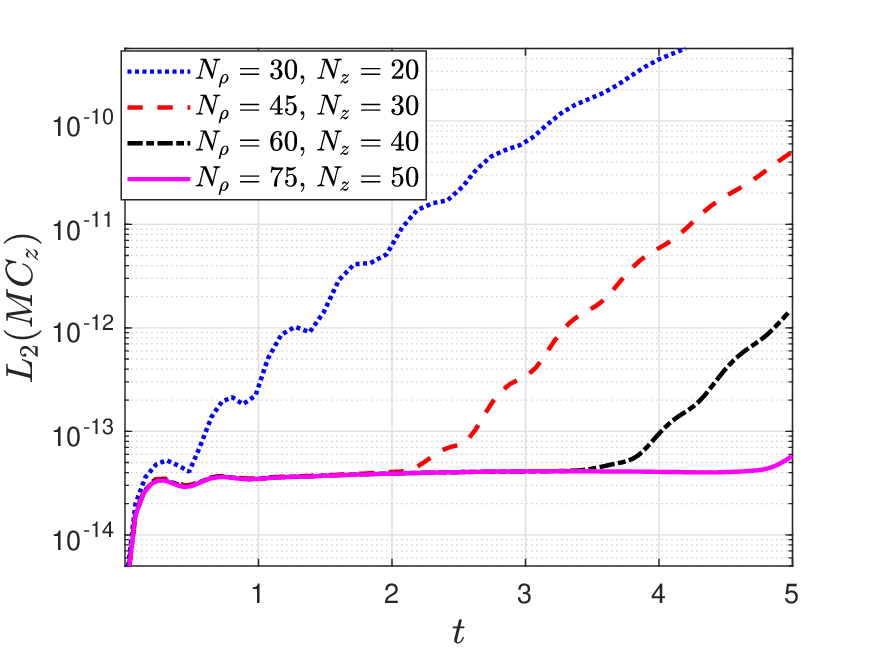}%{L2MC3_TW}
	\caption{Evolution of the violation of the momentum constraints $L_2(MC_\rho)$ and $L_2(MC_z)$ (upper and lower panels, respectively) for distinct resolutions, $N_{\rho}=30,45,60$ and $70$. The initial data is the axisymmetric Teukolsky solution with amplitude $A_0=10^{-7}$, and we have used the shock avoiding gauge with $\kappa_0=2$ and $L_{\rho}=L_z=5$.}
\end{figure}

\subsection{Weak gravitational waves: testing the Teukolsky exact solution}

%\begin{figure}[h]
%\includegraphics[width=7.5cm,height=5.5cm]{L2HC_tw_linear_k0_2}
%\caption{Evolution of the violation of the Hamiltonian constraint $L_2(HC)$ for distinct resolutions, $N_\rho=30,45$ and $60$. The initial data is the axisymmetric Teulkosky solution with amplitude $A_0=10^{-7}$. Here, we have used the shock avoiding gauge with $\kappa_0=2$ and $L_\rho=L_z=5$.}
%\end{figure}

%\begin{figure}[htb]
%\includegraphics[width=7.5cm,height=5.5cm]{L2MC1_tw_linear_k0_2}%{L2MC1_TW}\\
%\\
%\includegraphics[width=7.5cm,height=5.5cm]{L2MC3_tw_linear_k0_2}%{L2MC3_TW}
	%\caption{Evolution of the violation of the momentum constraints $L_2(MC_\rho)$ and $L_2(MC_z)$ (upper and lower panels, respectively) for distinct resolutions, $N_{\rho}=30,45,60$ and $70$. The initial data is the axisymmetric Teulkosky solution with amplitude $A_0=10^{-7}$, and we have used the shock avoiding gauge with $\kappa_0=2$ and $L_{\rho}=L_z=5$.}
%\end{figure}

The second test refers to the dynamics of weak gravitational waves on a flat Minkowski background provided by the Teukolsky solution with a seed function given by equation (\ref{seedteukolsky}). We use this solution as the initial data producing the following initial metric functions: 

\[\alpha=1, K=\bar{a}_{\rho \rho}=\bar{a}_{\theta \theta}=a_{\rho z}=\phi=0\]. 

\noindent The exact expressions for $\bar{h}_{\rho \rho},\,\bar{h}_{\theta \theta},\,h_{\rho z},\,h_{zz },\,\bar{\Lambda}^\rho,\,\bar{\Lambda}^z$ are generated by the exact Teukolsky solution translated into cylindrical coordinates. Moreover, we also considered the spectral code with the maximal slicing condition, where initially $\beta^\rho=\beta^z=0$. For the simulations, we assumed the same initial parameters used by Baumgarte {\it et al.} \cite{montero}, namely $\lambda=1$, $r_0 = 0$ and the small amplitude $A_0=10^{-7}$. 

\begin{figure}[htb]
\includegraphics[width=7.5cm,height=5.5cm]{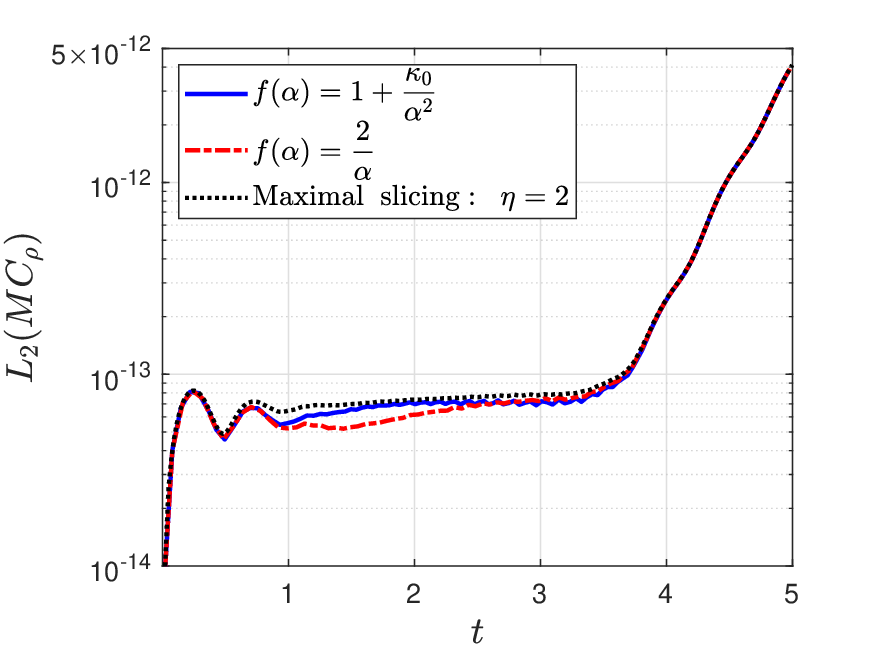}
\includegraphics[width=7.5cm,height=5.5cm]{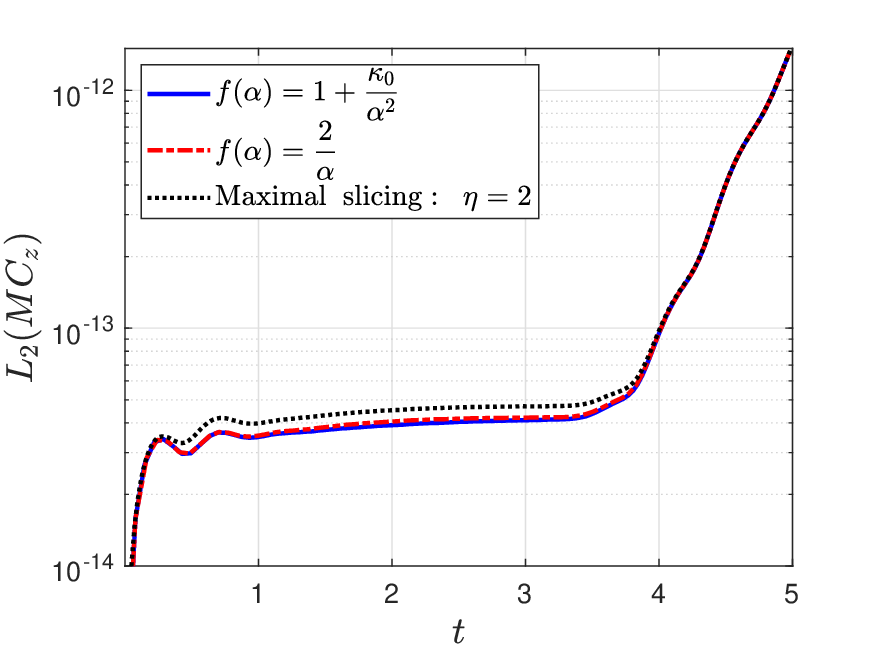}
	\caption{Comparison the evolution of $L_2(MC_{\rho})$ and $L_2(MC_z)$ (upper and lower panels, respectively) for distinct gauge conditions, namely the shock avoiding gauge (continuous line), the $1+log$ slicing (dash-dotted line), and the maximal slicing condition (dotted line) with  $N_\rho=60,\,N_z=40$. The initial data is the axisymmetric Teulkosky solution with amplitude $A_0=10^{-7}$ and $L_{\rho}=L_z=5$.}
\end{figure}

Instead of reconstructing the metric coefficients numerically and comparing them with the corresponding analytical solutions, we evolve the $L_2$-norms of the Hamiltonian and momentum constraints under distinct resolutions and gauge conditions. 

In the first experiment, we follow the evolution of $L_2(HC)$ using the shock avoiding slicing condition with  $\kappa_0=2$ for distinct truncations orders where $N_z=2 N_\rho /3$ and $N_\rho=30,45,60$. In Fig. 3, the results show a consistent decay of the Hamiltonian violation with increased numerical resolution. Notice that $L_2(HC)$ is about of the expected order of $\mathcal{O}(A_0^2) \sim 10^{-13}$ as expected. We have obtained similar results considering the {\it 1+log}, the harmonic, and the maximal slicing conditions.

\begin{figure}[htb]
\includegraphics[width=7.5cm,height=5.5cm]{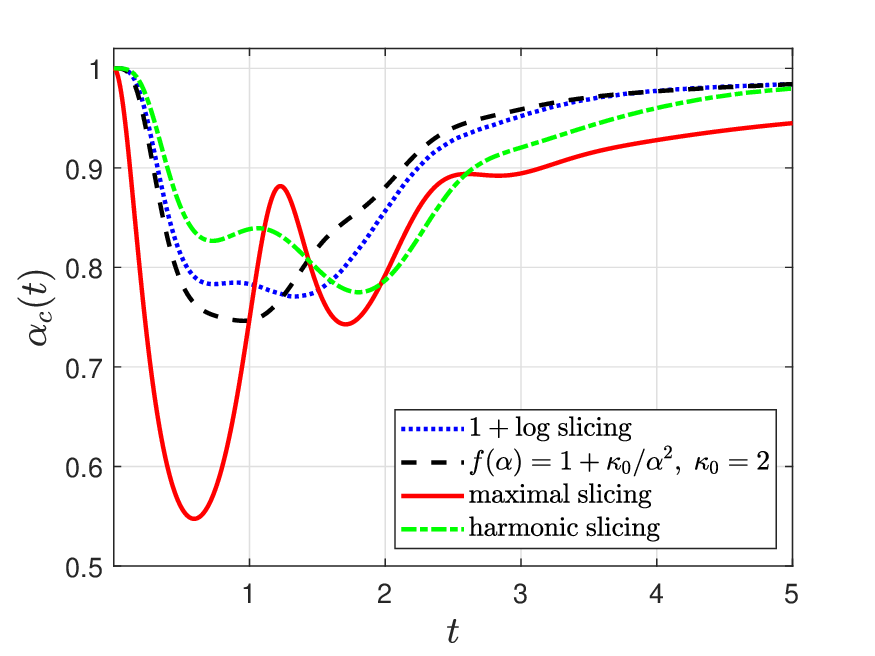}
	\caption{Evolution of the lapse at the origin under distinct gauge conditions for the Brill waves with $A_0=2.0$. We have used the resolution $N_{\rho}=90,\,N_z=60$ and map parameters $L_{\rho}=Lz=5$. }
\end{figure}

Turning to the momentum violation constraints, we show in Fig. 4 the evolution of $L_2(MC_\rho)$ and $L_2(MC_z)$ considering the same increase in the numerical resolution as of in Fig. 3, but including $N_\rho=75,\, N_z=50$. Next, in Fig. 5, we fix the resolution $N_\rho=60,\, N_z=40$ and evolve $L_2(MC_\rho)$ and $L_2(MC_z)$ under distinct gauge choices, namely, the shock avoiding slicing with $\kappa_0=2$, the {\it 1+log} slicing, and the maximal slicing gauges. All three gauge choices produce similar results regarding reproducing linear Teukolsky waves.

\subsection{Brill waves and Teukolsky waves: nonlinear evolution}

We now consider the nonlinear evolution of Brill waves. The seed function $q(\rho,z)$ we have adopted is given by equation (\ref{holz}) with $\lambda=1$, and we start choosing $A_0=2$ which guarantees the triggering of the nonlinear evolution of gravitational waves, although not strong enough to form an apparent horizon. After specifying the seed function, we obtain directly the initial metric coefficients $\bar{h}_{\rho \rho},\, \bar{h}_{\theta \theta}$, and $h_{zz}$. %The initial conformal factor $\phi(t_0,\rho,z)$ is determined by solving the Hamiltonian constraint (35), and the remaining functions vanish initially except the lapse, $\alpha(t_0,\rho,z)=1$.

The initial conformal factor $\phi(t_0,\rho,z)=\ln \psi(\rho,z)$ is determined by solving the Hamiltonian constraint (\ref{teukolskycons}). To this aim, it is necessary to establish an appropriate spectral approximation for $\psi(\rho,z)$, that is

\begin{equation}
\psi(\rho,z)=1+\sum_{k=0}^{N_\rho}\sum_{j=0}^{N_z}\,\hat{\psi}_{kj}(t)SB_{2k}(\rho)SB_{2j}(z).
\end{equation}

\noindent Therefore, by imposing the residual equation associated with the Hamiltonian constraint vanishes at the grid points $(\rho_l,z_m)$, $l=0,1,.., N_\rho,\;m=0,1,.., N_z$, we obtain a set of linear algebraic equations for the coefficients $\hat{\psi}_{kj}$ that fixes the conformal factor $\psi(\rho,z)$ and consequently $\phi(t_0,\rho,z)$. The remaining BSSN functions vanish initially (including $\beta^\rho=\beta^z=0$ for the maximal slicing evolution) except the lapse, $\alpha(t_0,\rho,z)=1$.

%\begin{figure}[htb]
%\includegraphics[width=7.5cm,height=5.5cm]{fig_lapse_gauges}
%	\caption{Evolution of the lapse under distinct gauge conditions for the Brill waves with $A0=2.0$. We have used the resolution $N_{\rho}=90,\,N_z=60$ and map parameters $L_{\rho}=Lz=5$. }
%\end{figure}

\begin{figure}[htb]
\includegraphics[width=7.5cm,height=5.5cm]{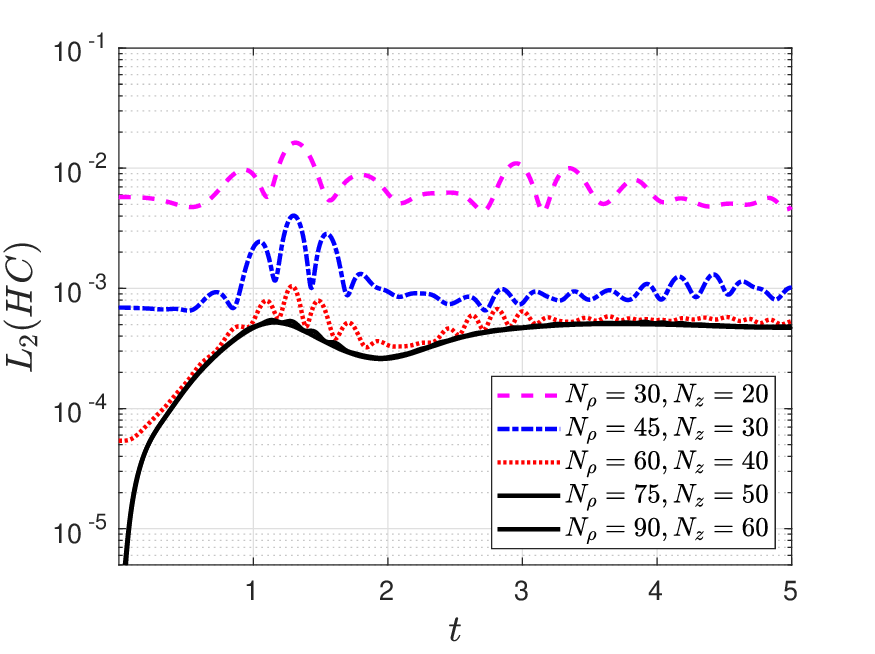}
\includegraphics[width=7.5cm,height=5.5cm]{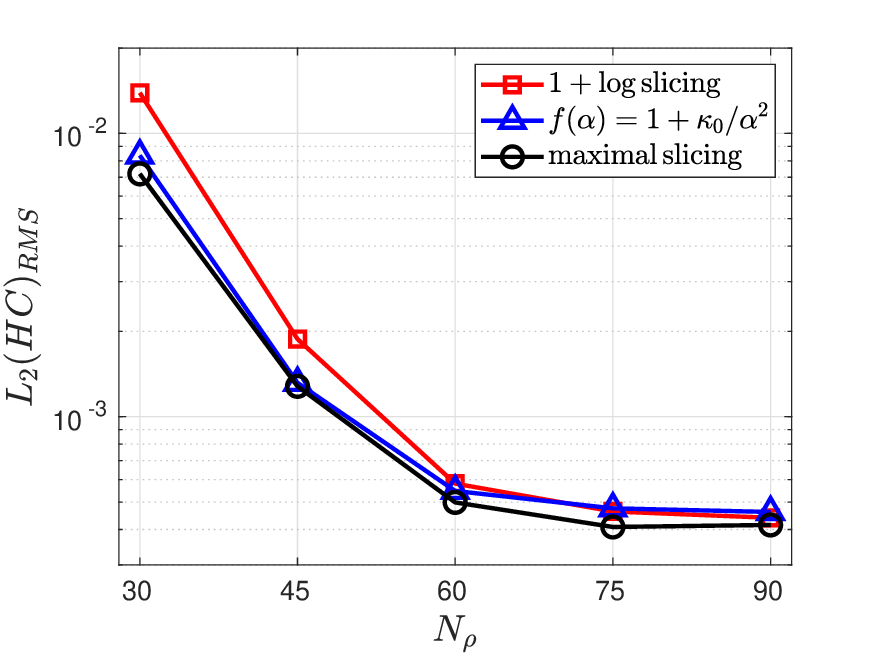}
	\caption{Upper panel: evolution of the Hamiltonian constraint violation for distinct resolutions and using the maximal slicing condition. Lower panel: convergence of the RMS value of the Hamiltonian constraint violation for the $1+log$ slicing, the shock avoiding slicing 
 ($\kappa_0=2)$, and the maximal slicing conditions.}
\end{figure}

We focus initially on the lapse behavior at the origin, $\alpha_c(t)=\alpha(t,r=0)$, considering several gauge choices. As an illustration, we present in Fig. 6 the evolution of $\alpha_c(t)$ for the {\it 1+log} slicing, the shock-avoiding slicing, harmonic slicing, and the maximal slicing. We have used the map parameters $L_{\rho}=L_z=5$ and truncation orders $N_{\rho}=90, N_z=60$; for the shock-avoiding slicing, $\kappa_0=2$, and $\eta=6$ for the Gamma-driver condition in the maximal slicing gauge. In all cases, despite being structurally distinct,  $\alpha_c(t)$ tends asymptotically to unity, indicating total dispersion of the collapsing wave package. Regarding the Hamiltonian violation, Fig. 7(a) shows the evolution of $L_2(HC)$ for distinct resolutions and using the maximal slicing gauge. We have noticed that the error saturates about $10^{-4})$ for the resolution $N_\rho=75, N_z=50$. Moreover, we have obtained similar results with other gauge conditions, as shown in Fig. 7(b) with the decay of the RMS value of $L_2(HC)$ generated by the {\it 1+log} slicing, the shock avoiding slicing, and the maximal slicing gauges. 

Next, we have considered the initial data constructed by the Teukolsky exact solution. As in the last subsection, we initially fix the metric coefficients $\bar{h}_{\rho \rho},\,\bar{h}_{\theta \theta},\,h_{z z}$, and $h_{\rho z}$ with the exact corresponding expressions taking $\lambda=1/2$ and $r_0=2$ (cf. \cite{hild_13}). We obtain the initial conformal factor $\phi(0,\rho,z)$ as described for the Brill waves initial data, i.e., after solving the Hamiltonian constraint (34) converted in cylindrical coordinates for the function $\psi(\rho,z)$ and then $\phi(t_0,\rho,z)=\ln \psi(\rho,z)$ following the same procedure delineated for the Brill wave initial data.

\begin{figure}[htb]
\includegraphics[width=7.5cm,height=5.5cm]{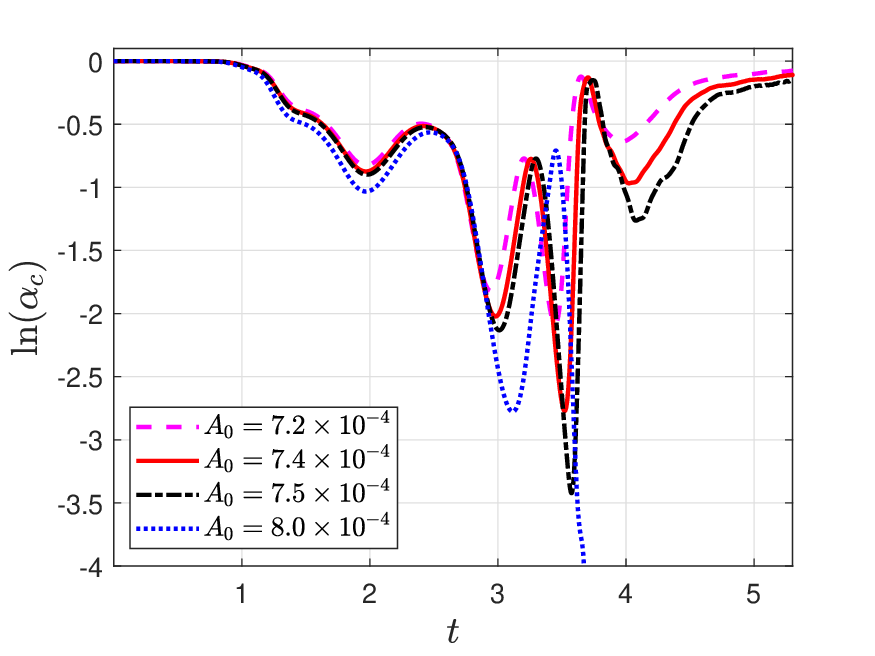}
	\caption{Evolution of the logarithm of the lapse evaluated at the origin using the $1+log$ slicing condition for the Teukolsky initial data considering distinct amplitudes as indicated. Notice that $A_0=8.0 \times 10^{-4}$ signalizes the formation of an apparent horizon. In most simulations, we have the resolutions $N_\rho=140$, $N_z=70$, and the map parameters $L\rho=1.0\, L_z=2.0$. For the highest amplitudes, we set the resolution $N_\rho=160$, $N_z=80$.}
\end{figure}

\begin{figure}[htb]
\includegraphics[width=7.5cm,height=5.5cm]{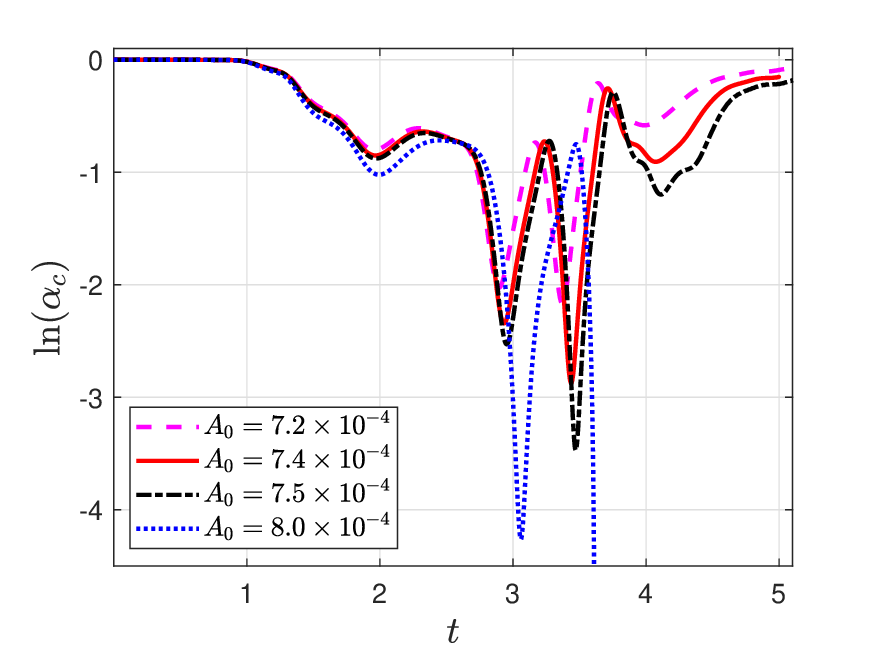}
	\caption{Evolution of the logarithm of the lapse evaluated at the origin with the shock avoiding slicing condition with $\kappa_0=2/3$ for the Teukolsky initial data considering the same amplitudes of Fig. 8. Notice that the plots are similar to those of Fig. 8 despite using distinct gauge conditions. Also, $A_0=8.0 \times 10^{-4}$ signalizes the formation of an apparent horizon.}
\end{figure}

The most favorable gauges for evolving the Teukolsky waves are the $1+log$ and the shock avoiding gauges. In Figs. 8 and 9, we have shown the evolution of the lapse at the origin for the initial amplitudes $A_0=7.2 \times 10^{-4},\,7.4 \times 10^{-4},\,7.5 \times 10^{-4}$ and $8.0 \times 10^{-4}$. In both plots, the formation of an apparent horizon takes place for $A_0=8.0 \times 10^{-4}$, which is consistent with Ref. \cite{hild_13}. that signalizes the critical amplitude about $1.5 \time 10^{-3}$ since our initial amplitude is twice of theirs. 

In particular, for the $1+log$ slicing condition, we have noticed that the lapse at the origin profile resulting from $A_0=7.4 \times 10^{-4}$ is equivalent to the resulting profile of Ref. \cite{hild_13} with $A=1.44 \times 10^{-3}$, as should be expected. Another aspect worth mentioning is the similarity between the plots resulting from the $1+log$ slicing with the correspondent obtained with the shock avoiding gauge where we have set $\kappa_0=2/3$. If $\kappa_0=2$ as employed previously, the evolution crashes for some of the amplitudes near the critical one.

\begin{figure}[htb]
\includegraphics[width=7.5cm,height=5.5cm]{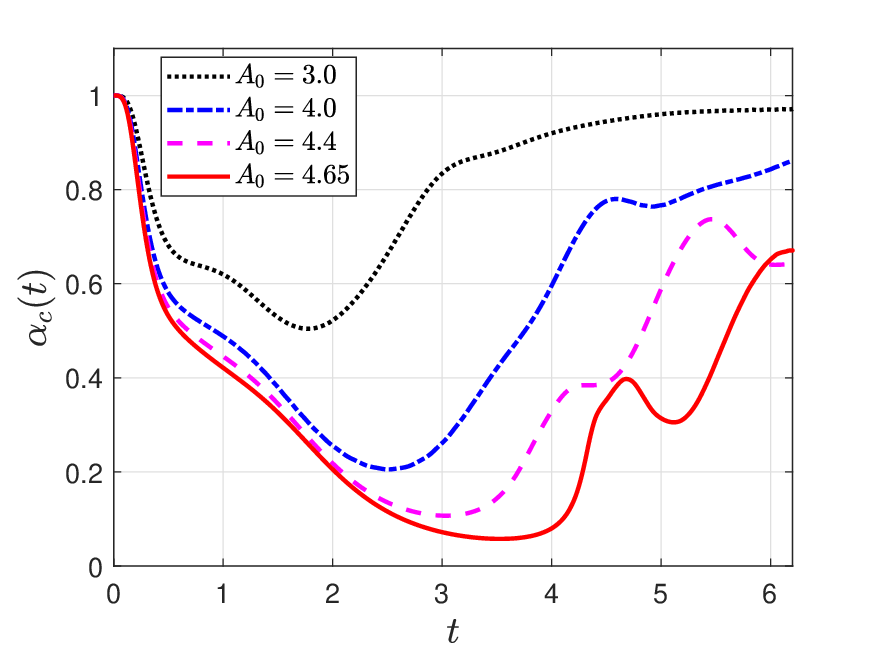}
	\caption{Evolution of the lapse evaluated at the origin with the $1+log$ slicing condition for high amplitudes Brill waves denoted by $A_0=3.0, 3.0, 4.4$ and $4.65$.}
\end{figure}

The last numerical experiment consisted of evolving high amplitudes Brill waves. We summarized the results in Fig. 10 with the plots of the lapse evaluated at the origin for the initial amplitudes $A_0=3.0, 4.0, 4.4$ and $4.65$ taking $\lambda=1$ and $\rho_0=0$ In Eq. (39). In these simulations, we have set $L_\rho=1$ and $L_z=3$ for the map parameters, and the maximal resolution $N_\rho=240$, $N_z=80$. We have adopted the $1+log$ slicing conditions which showed to be the most efficient gauge condition in evolving high amplitudes of the Brill waves. For instance, using Alcubierre's slicing condition (56), the integration fails for $A_0 \geq 4.5$, but a better understanding is necessary in order to explain this feature.

\section{Final Remarks}
\label{sec06}
%
%\indent \par In this work, we present a numerical code which integrates the equations of General Relativity, according to the BSSN formalism in cylindrical coordinates. One of the main features of the code is the fact that we use pseudospectral techniques, whose basis obeys the regularity conditions at the origin and infinity of the cylindrical coordinate system. The main methodology for testing numerical codes relies on the use of two General Relativity solutions that describe gravitational waves: Teukolsky and Brill. As a first test, even before using these gravitational wave solutions, we analyzed the evolution of initial pure gauge data. The results focus on measuring the violation of the Hamiltonian momentum bonds and these were addressed under different gauge conditions. The main calibers used were $1+log$, harmonic and Alcubierre shock avoiding. For the maximum numerical resolution adopted, the minimum violations of the Hamiltonian bond (for pure gauge evolution) are of the order of $10^{-8}$ for the shock avoiding and $1+log$ gauges. While the violations for the momentum link are on the order of $10^{-5}$ the same gauge choices. It was verified that, for the momentum link, the errors already saturate at lower resolutions of the numerical mesh, which remains constant for higher resolutions.
%

In this work, we introduce a robust numerical code that seamlessly integrates the equations of General Relativity according to the BSSN formalism in cylindrical coordinates. A key feature of our code is using pseudospectral techniques, specifically the Galerkin-Collocation method, whose basis adheres to the cylindrical coordinate system's regularity conditions at the origin and infinity.

%\indent \par The second test was through the use of Teukolsky's solution for weak gravitational waves that are characterized by low amplitudes.  For this regime, amplitudes of the order of $A_0=10^{-7}$ were considered. We also analyze violations of the Hamiltonian and momentum constraints for different resolutions, as well as their evolution over time. In the case of maximum resolution, violations were around $10^{-13}$, that is, of the order of $A_{0}^{2}$. At the end of the second test, we also verified that the constraint violations do not vary considerably when changing the gauge conditions.
%
The principal methodology for testing our numerical code relies on standard convergence and accuracy tests under distinct gauge choices. We have considered the evolution of pure gauge data, verified the Teukolsky solution, which is valid in the linear regime, and evolved the Brill and the Teukolsky initial data waves in the nonlinear regime. The results focus on measuring the violation of the Hamiltonian and the momentum constraint addressed under different gauge conditions. In this vein, the slicing conditions used are: the $1+log$ slicing condition, the harmonic gauge, the Alcubierre shock-avoiding slicing condition, the maximal slicing condition.

%\indent \par The third test took into account the evolution of the BSSN code from Brill and Teukolsky solutions with larger amplitudes when compared with the second test cases. Brill's solution is, by construction, a nonlinear solution of General Relativity, which is not the case for the Teukolsky quadrupole solution. However, in the evolution of the numerical code, we characterize the increase in nonlinearity through the amplitude parameter $A_0$ contained in each of the seed functions. Both Brill and Teukolsky descriptions, with larger amplitudes ($A_0 \sim 10^{-3}$), presented violations of the Hamiltonian constraint of the same order for all different gauge choices at higher resolutions of the numerical grid. Over time, violations saturate at around $10^{-3}$ for the maximum resolution adopted.
%

We have noticed that the minimum violations of the Hamiltonian constraint for pure gauge evolution are about $10^{-8}$ for the $1+log$ and the shock-avoiding gauges at higher resolutions. At the same time, the violations for the momentum constraint saturate at lower resolutions about $10^{-5}$ but are slightly better for the $1+log$ and shock-avoiding gauges. In this experiment, the harmonic gauge produced somewhat less favorable results.

The second test used Teukolsky's solution for weak gravitational waves characterized by low amplitudes. We set the initial amplitude for this regime as $A_0=10^{-7}$. We also analyzed violations of the Hamiltonian and momentum constraints for different resolutions and their evolution over time. In the case of maximum resolution, violations were around $ 10^ {-13} $, that is, of the order of $A_{0}^{2}$. At the end of the second test, we verified that the constraint violations are approximately independent of the gauge choices, demonstrating the versatility and robustness of our numerical code. In this experiment, we have considered the maximal slicing, the $1+log$ slicing condition, and the shock-avoiding slicing.

%\indent \par In addition to the constraint violations, in the third test, we verified the behavior of the lapse function for different resolutions and gauge conditions. The behavior of the lapse function proved to be stable and in accordance with other results in the literature, which motivates us to analyze the numerical evolution based on other initial data with axial symmetry and in cylindrical coordinates. As future work proposals, we can point out the following ideas: an analysis of the behavior of the ADM mass for all tests considered here in this work; obtaining the gravitational wave modes for Brill and Teukolsky evolutions and verifying the behavior of the angular and temporal patterns of such waves; consider other solutions of General Relativity that have axial symmetry in the presence of matter terms.

The third test considered the evolution of the BSSN code from Brill and Teukolsky solutions with larger amplitudes than the second test cases. Brill's solution is, by construction, a nonlinear solution of General Relativity, which is not the case for the Teukolsky quadrupole solution. 

However, in the evolution of the numerical code, we characterize the increase in nonlinearity through the amplitude parameter $A_0$ contained in each seed function. With larger initial amplitudes, the gauge choices maximal slicing, $1+log$ slicing, and shock-avoiding slicing produced similar results regarding the decay of the Hamiltonian constraint. The harmonic gauge was the most unfavorable in this experiment. We highlight that one of the virtues of the present code is the effectiveness of relatively low resolutions adopted. For instance, in this nonlinear experiment, the maximal resolution was $N_\rho=90,\, N_z=60$, or $5,551$ collocation points. 

In the last part of the nonlinear evolutions, we explored the code's performance by increasing the initial amplitudes so that apparent horizons are eventually formed. We have obtained satisfactory results with the Teukolsky waves, where $A_0=8.0 \times 10^{-4}$ signalizes the formation of an apparent horizon, which agrees with the simulations of Ref. \cite{hild_13}. The profiles of the lapse evaluated at the center $(\rho=0,z=0)$ are similar for the $1+log$ slicing and shock-avoiding slicing with $\kappa_0=2/3$. Both gauges produced the best results. 

In trying to reach the critical amplitudes for the Brill waves, we encountered some difficulties in the simulations with amplitudes greater than $A_0=4.65$. The maximal resolution was $N_\rho=240,\, N_z=40$ or $19,521$ collocation points, in contrast with the maximal resolutions for the nonlinear Teukolsky waves $N_\rho=140,\, N_z=70$ or $10,011$ collocation points.

As future work proposals, we can point out the following ideas: possible scaling relations involving the ADM mass, the determination of the gravitational wave modes for Brill and Teukolsky evolutions, and verifying the behavior of the angular and temporal patterns of such waves, and the investigation of other solutions of General Relativity that have axial symmetry in the presence of matter terms.

\bibliography{apssamp}% Produces the bibliography via BibTeX.

\appendix

\section{Obtaining the explicit form of the BSSN equations}
\label{app}
\indent \par In this appendix we will describe the methodology of explicitly obtaining the generalized BSSN equations in cylindrical coordinates. It is important to note that we use the package for {\it{Maple}}, {\it{GRTensorIII}}, through its repository in {\it{GitHub}} \cite{github}.
\indent \par As an example, we use equation (\ref{eqK}) associated with the trace of the extrinsic curvature, $K$. Basically the idea is to use the function \texttt{grdef} in order to write the equations as similar as possible to the equations in the notation of Section \ref{sec02}. In addition, it is also interesting to divide the equations into smaller parts, so that their transition to computational notation is cleaner. In this example we split the equation for $K$ into 3 parts, two containing the halves of the equation and the third being represented by the Laplacian of $\alpha$,\\

\noindent \texttt{grdef(`dtK1\{ \} := (alpha/3) * K\textasciicircum 2 + alpha * A\{i j\} * A\{k l\} * g\{\textasciicircum k \textasciicircum i{\}} * g{\{}\textasciicircum l \textasciicircum j\}`);} \\
\noindent \texttt{grdef(`dtK2\{ \} := - exp(-4*phi) * (Lapalpha + 2 * g\{\textasciicircum i \textasciicircum j\} * Dalpha\{j\} * Dphi\{i\})`);} 

\vspace{2mm}
\noindent and
\vspace{2mm}

\noindent \texttt{grdef(`dtK\{ \} := dtK1\{ \} + dtK2\{ \}`);}\\

\noindent In the above, \texttt{dtK\{\}}, \texttt{alpha}, \texttt{A\{i j\}}, \texttt{g\{\textasciicircum k \textasciicircum i{\}}}, \texttt{phi}, \texttt{Lapalpha}, \texttt{Dalpha\{j\}} and \texttt{Dphi\{i\}} represent $\partial_t K$, $\alpha$, $\tilde{A}_{ij}$, $\bar{\gamma}^{k i}$, $\phi$, $\bar{\nabla}^2 \alpha$, $\bar{\nabla}_{j} \alpha$ and $\bar{\nabla}_{i} \phi$ respectively.\\
\noindent The result of the above definitions generates the following equation,

\begin{eqnarray}
\label{a1}
\nonumber
\partial_t K &=& \frac{1}{3}\alpha K^{2}+\frac{\alpha}{\Delta^{2}}(h_{zz}^{2}a_{\rho \rho}^{2} - 4h_{\rho z}h_{zz}a_{\rho z}a_{\rho \rho} + 2 h_{\rho z}^{2}a_{zz}a_{\rho \rho} + \\
\nonumber
&+& 2 h_{\rho \rho}h_{zz}a_{\rho z}^{2} + 2 h_{\rho z}^{2}a_{\rho z}^{2} - 
4h_{\rho \rho}h_{\rho z}a_{zz}a_{\rho z} + \frac{(a_{\theta \theta}\Delta)^{2} }{h_{\theta \theta}^{2}} +\\
\nonumber
&+& h_{\rho \rho}^{2}a_{zz}^{2}) - e^{-4\phi}[\Delta^{-1}(2 h_{zz} \partial_{\rho} \alpha \partial_{\rho}\phi - 2h_{\rho z}\partial_{z}\alpha \partial_{\rho} \phi- \\
&-& 2h_{\rho z}\partial_{\rho}\alpha \partial_{z}\phi + 2h_{\rho \rho}\partial_{z}\alpha \partial_{z} \phi) + \nabla^{2}\alpha], 
\end{eqnarray}
\noindent where,
\begin{eqnarray}
\Delta \equiv h_{\rho \rho}h_{zz}-h_{\rho z}^2.\\
\nonumber
\end{eqnarray}
\noindent The expression for $\bar{\nabla}^2 \alpha$ depends on the metric components exclusively. So, once the metric functions are determined for each time $t$, the laplacian is uniquely obtained and substituted in equation (\ref{a1}). For the other equations we do the same procedure, just redefining the remaining BSSN variables in the {\it{GRtensorIII}} notation.

\end{document}